\documentclass{article}
\usepackage{spconf,amsmath,graphicx}

\usepackage[per-mode=symbol]{siunitx}
\usepackage{tikz, pgfplots}
\pgfplotsset{compat=1.15}
\usepgfplotslibrary{groupplots}
\usepackage{subcaption}
\usepackage{multirow}
\usepackage{booktabs}
\usepackage{url}

\newcommand{\sd}[1]{\textcolor{black}{#1}}
\newcommand{\lb}[1]{\textcolor{black}{#1}}

\title{Can Synthetic Data Boost the Training of Deep Acoustic Vehicle Counting Networks?}

\name{Stefano Damiano$^{1*}$, Luca Bondi$^2$, Shabnam Ghaffarzadegan$^2$, Andre Guntoro$^2$, Toon van Waterschoot$^{1}$\sthanks{This project has received funding from the European Union's Horizon 2020 research and innovation programme under the Marie Sk\l{}odowska-Curie grant agreement No. 956962 and from the European Research Council under the European Union's Horizon 2020 research and innovation program / ERC Consolidator Grant: SONORA (no. 773268). This paper reflects only the authors' views and the Union is not liable for any use that may be made of the contained information. \endgraf \copyright  2024 IEEE. Personal use of this material is permitted. Permission from IEEE must be obtained for all other uses, in any current or future media, including reprinting/republishing this material for advertising or promotional purposes, creating new collective works, for resale or redistribution to servers or lists, or reuse of any copyrighted component of this work in other works.}}
\address{$^1$ Dept. of Electrical Engineering (ESAT-STADIUS), KU Leuven, 3001 Leuven, Belgium\\
	$^2$ Bosch Research, Bosch Center for Artificial Intelligence
  }
\begin{document}
\maketitle
\begin{abstract}
In the design of traffic monitoring solutions for optimizing the urban mobility infrastructure, acoustic vehicle counting models have received attention due to their cost effectiveness and energy efficiency.
Although deep learning has proven effective for visual traffic monitoring, its use has not been thoroughly investigated in the audio domain, likely due to real-world data scarcity. 
In this work, we propose a novel approach to acoustic vehicle counting by developing: i) a traffic noise simulation framework to synthesize realistic vehicle pass-by events; ii) a strategy to mix synthetic and real data to train a deep-learning model for traffic counting. The proposed system is capable of simultaneously counting cars and commercial vehicles driving on a two-lane road, and identifying their direction of travel under moderate traffic density conditions. 
With only 24 hours of labeled real-world traffic noise, we are able to improve counting accuracy on real-world data from $63\%$ to $88\%$ for cars and from $86\%$ to $94\%$ for commercial vehicles.

\end{abstract}
\section{Introduction}
\label{sec:introduction}

The development of smart cities relies on the deployment of sensors and devices in urban areas to collect data and efficiently monitor and manage public infrastructures~\cite{du_sensable_2019}. Traffic monitoring systems are used to gather information on the usage of roadways, including estimates on the number, speed and type of vehicles passing by, to control the traffic volume or to detect anomalous conditions. Several different sensors are available for traffic monitoring, including intrusive systems embedded in the road, e.g. induction loops, vibration or magnetic sensors, non-intrusive systems mounted over or on the side of the road, e.g. radar, cameras, infrared or acoustic sensors, and off-road mobile devices, e.g. aircraft or satellites~\cite{won_intelligent_2020, djukanovic_robust_2020}. Acoustic sensors, though not being the most common solution, provide several advantages compared to other devices: microphones are in fact low-cost and power efficient devices, not affected by low-visibility conditions.

Although the literature is still relatively limited, several approaches for acoustic vehicle detection~\cite{ishida_saved:_2018,wang_real-time_2022,szwoch_acoustic_2021,kubo_design_2018,bulatovic_mel-spectrogram_2022} and counting~\cite{djukanovic_robust_2020,djukanovic_neural_2021,zu_vehicle_2017,severdaks_vehicle_2013} have been proposed. These methods are mostly based on the analysis of audio signals captured using either a single microphone or a microphone array, rely either on traditional signal processing~\cite{djukanovic_robust_2020}, deep learning~\cite{wang_real-time_2022} or a combination of both~\cite{djukanovic_neural_2021}, and target multiple tasks. While the basic goal is the detection of passing vehicles, more advanced methods also aim at discriminating the type of vehicle (e.g. car, truck, motorbike, etc.), the direction of motion (e.g. left-to-right or right-to-left), and, eventually, its speed. 
Even though data-driven approaches prove effective in sound detection tasks~\cite{mesaros_sound_2021}, their potential has not yet been thoroughly investigated in the context of acoustic vehicle counting (AVC). One reason can be identified in the scarcity of available datasets for this purpose: a few existing datasets~\cite{djukanovic_robust_2020, abesser_IDMT-traffic:_2021} are of limited size, insufficient for training an end-to-end deep learning model, and usually contain single-channel recordings.
Collecting data is in fact a complex and expensive task that entails not only recording audio, but also collecting ground truth data using other sensor modalities, usually difficult to deploy, and developing synchronization strategies to align the collected audio and ground truth sensor(s) data.

\begin{figure*}[t]
  \centering
  \resizebox{0.95\linewidth}{!}{
    \includegraphics{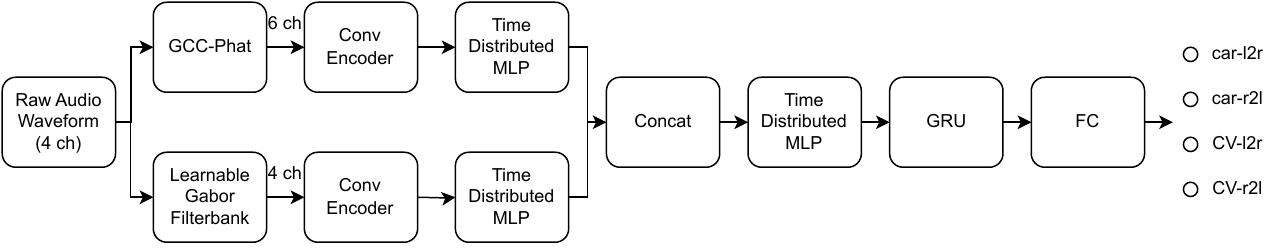}
  }
  \caption{The proposed CRNN architecture takes as input the raw signal from a 4-channel linear microphone array and computes in parallel: i) the Generalized Cross-Correlation with Phase Transform (GCC-Phat) between pairs of channels; ii) a Learnable Filterbank with Gabor filters. Two convolutional encoders followed by Time-Distributed Multi-Layer Perceptrons (TD-MLP) compute spatial and semantic features, respectively. The concatenated features are processed by a further TD-MLP layer, followed by a Gated Recurrent Unit (GRU) and a fully connected (FC) layer to regress the number of vehicles per type (car, CV) and per direction (left-to-right, right-to-left).}
  \vspace{-5pt}
  \label{fig:CRNN_architecture}
\end{figure*}
\begin{figure*}[t]
\centering
  \includegraphics{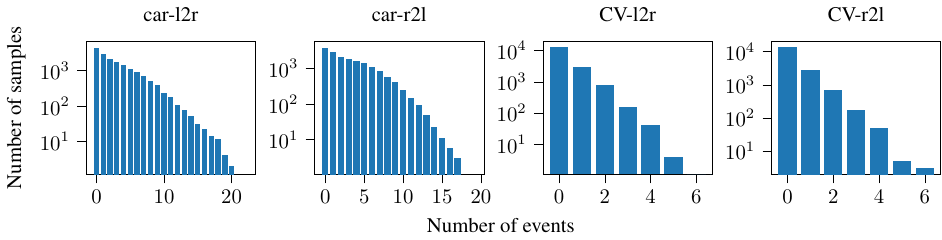}
  \caption{Distribution of recorded events in four target categories averaged over $\SI{60}{\second}$ audio segments.}
  \label{fig:data_distribution}
  \vspace{-5pt}
\end{figure*}

In this paper we address AVC using a 4-channel linear microphone array deployed on the side of a two-lane road. Our goal is to identify pass-by events in moderate traffic density conditions - i.e. up to 1000 vehicles per hour, per lane - with vehicles driving from left to right in the close lane and in the opposite direction in the far lane. For each detected event, the vehicle must be classified as \emph{car} or \emph{commercial vehicle} (CV) - heavy vehicles such as trucks and buses - and the direction of transit must be identified. 
In this work, we introduce a convolutional recurrent neural network (CRNN) for traffic counting, with a training procedure based on the usage of synthetic data generated using the \emph{pyroadacoustics} simulator~\cite{damiano_pyroadacoustics:_2022}. This method  drastically reduces the amount of required real-world data to achieve high traffic counting accuracy. In summary, we first define a strategy to synthesize acoustic traffic noise, then pre-train the model based on the synthetically generated dataset, and finally perform fine-tuning on a limited amount of real-world data. Evaluation on a separate dataset collected by Bosch on a German municipal road shows that the proposed system is capable of learning to count traffic effectively with as little as $\SI{24}{\hour}$ of labeled real-world data. \sd{Source code for this paper is available online\footnote{[Coming soon] \url{https://github.com/boschresearch/acoustic-traffic-simulation-counting}}}.

\section{Methodology}
\label{sec:proposed_method}
\subsection{Data Description}
\label{subsec:dataset_description}
A traffic noise dataset is provided by Bosch for the design and assessment of the AVC model. Data was recorded in the span of 11 days (264 total hours of recording) over few months, on a two-lane municipal road with a fixed recording setup of a 4-channel linear microphone array. The array is located at the side of the road, with parallel orientation with respect to the traffic direction, at a lateral distance of $\SI{4}{\meter}$ and a height of $\SI{2.7}{\meter}$. The microphones are uniformly spaced, with an array aperture of $\SI{24}{\centi\meter}$. The road has two lanes in which vehicles travel in opposite directions, passing from left to right (l2r) in the closest lane, and from right to left (r2l) in the farthest. The traffic flow has a maximum density of 1000 vehicles per hour, per lane, and includes various vehicle types such as: cars, commercial vehicles (i.e. buses and trucks), motorbikes and bicycles; with a maximum speed of $\SI{100}{\km\per\hour}$. \sd{Overlapping pass-bys (in different lanes) occur in the dataset}.
Ground truth data containing pass-by instant, vehicle type and direction of motion, are gathered via induction coils for cars and CVs, that together represent the majority of road agents and thus have the strongest impact on the traffic distribution.
Fig.~\ref{fig:data_distribution} shows the data distribution across the dataset for each of the four categories of interest (car-l2r, car-r2l, CV-l2r, CV-r2l).

\subsection{CRNN Architecture}
\label{subsec:CRNN_architecture}

To identify the number of passing vehicles in these four categories, we use the convolutional recurrent neural network (CRNN) depicted in Fig.~\ref{fig:CRNN_architecture}. In the upper network branch, Generalized Cross-Correlation with Phase transform (GCC-Phat) features, \sd{helpful to identify the direction of movement of the detected vehicles}, are provided as input to a convolutional encoder, that consists of two Conv2D layers with 32 filters each, followed by a Conv2D layer with 64 filters, each with kernel size (3,3) and a stride of 2 in both dimensions.
\begin{figure}[tb]
  \includegraphics{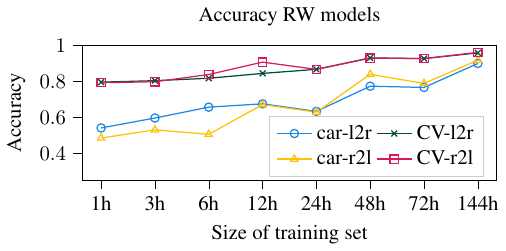}
  \vspace{-15pt}
  \caption{Test accuracy of models trained on an increasing amount of real-world data ($\text{RW}$).}
  \label{fig:res_train_real}
\end{figure}
\begin{figure}[tb]
  \vspace{-6pt}
  \includegraphics{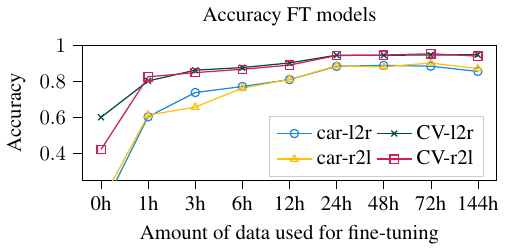}
  \vspace{-15pt}
  \caption{Test accuracy of models pre-trained on synthetic data and then fine-tuned ($\text{FT}$) on an increasing amount of real-world data.}
  \vspace{-5pt}
  \label{fig:acc_train_synth}
\end{figure}
\begin{figure}[tb]
  \includegraphics{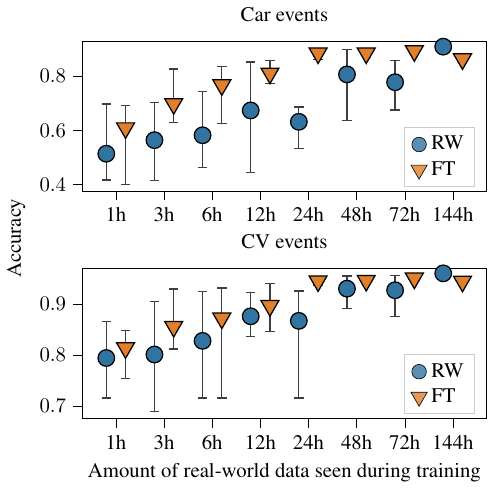}
  \caption{Average accuracy (marks) and accuracy ranges (whiskers) obtained using a model pre-trained on synthetic data and fine-tuned (FT) on an increasing amount of real-world data, and a model trained from scratch using same amount of real-world data (RW).}
  \label{fig:acc_comparison}
\end{figure}
The output of the encoder is then passed through two time-distributed fully connected (FC) layers with 128 neurons each. Time-distributed layers are independently applied to each time frame of their input. In the lower branch, the spectrogram of the 4-channel audio signal\sd{, useful to identify the type of vehicle (car or CV),} is filtered via a learnable Gabor filterbank~\cite{zeghidour_leaf:_2021} with 96 channels, and then passed through a convolutional encoder and a time-distributed block similar to the upper branch. The output of the two branches is then concatenated and passed to another time-distributed block with three layers of 128 neurons each. Finally, a gated recurrent unit (GRU) with two layers of 128 neurons is used to model temporal dependencies. A FC layer with four neurons is then used as a regression output, providing the count of vehicles of the four categories of car-l2r, car-r2l, CV-l2r, CV-r2l. The model has a total of $\SI{506}{\K}$ trainable parameters. %

\subsection{Synthetic Data Generation}
\label{subsec:data_generation}
The main challenge of deep learning models is the large amount of data required to train them. To ease this requirement, we propose a training strategy based on the use of synthetic data. The data generation system relies on the open source\footnote{\url{https://github.com/steDamiano/pyroadacoustics} }~\emph{pyroadacoustics} simulator~\cite{damiano_pyroadacoustics:_2022}, %
designed to simulate the audio produced by a source moving on a road - including direct sound, reflection from the asphalt, air absorption and Doppler effect - and received by a set of static omnidirectional microphones with arbitrary array geometry. %
First, we simulate individual pass-bys defining a rectilinear source trajectory, based on speed and direction of travel, to obtain a $\SI{30}{\second}$-long simulation. 
Since the simulation emulates the real-world dataset, and there are no traffic lights, crossroads or turns close to the array location, the speed of the vehicles is assumed to be constant during the pass-by for simulation purpose.
Then, we synthetically generate source signals according to the Harmonoise model~\cite{jonasson_acoustical_2007}, which models a passing vehicle by superimposing two vertically-stacked point sources, containing a mixture of road/tire interaction noise and engine noise. 
The lower source (LS), located at a height of $\SI{1}{\centi\meter}$, emits $80\%$ of the total power of the road/tire interaction noise, and $20\%$ of the total power of engine noise. The higher source (HS), located at a height of $\SI{30}{\centi\meter}$ for cars, and $\SI{75}{\centi\meter}$ for commercial vehicles, emits $20\%$ of the total power of the road/tire interaction noise, and $80\%$ of the total power of engine noise. 
The road/tire interaction noise signal is generated according to the Harmonoise model~\cite{jonasson_acoustical_2007}, while the engine noise is produced using the Baldan model~\cite{baldan_physically_2015}, and normalized to match the Harmonoise signal power as described in~\cite{fu_auralisation_2021}. After creating the signal mixtures of the HS and LS, we use them as input to \emph{pyroadacoustics} to simulate the pass-by events. 
Finally, the simulated individual events are combined based on the traffic distribution of the real-world data to re-create the traffic flow, and the resulting signal is split into $\SI{60}{\second}$-long audio segments. 

\section{Experimental Evaluation}
\label{sec:experiments}
In this section, we analyze the performance of the proposed model and the effectiveness of synthetic data for pre-training purpose.
We resample all the data to $\SI{16}{\kilo\Hz}$ and apply peak normalization to each audio segment. During training, we introduce spectrogram augmentation~\cite{park_specaugment:_2019}, use a mean squared error loss, Adam optimizer~\cite{kingma_adam:_2015} with learning rate $\mathrm{lr}=10^{-4}$, batch size of 128. We train the model for 150 epochs, selecting the best checkpoint based on validation loss. We implement the proposed architecture in Keras\footnote{\url{https://github.com/keras-team/keras}} and train on a single V100-32GB GPU. To compare and evaluate different training strategies, we first round the output of the CRNN to the nearest integer, then compute the classification accuracy, i.e. the number of segments where the predicted number of vehicles matches exactly the ground truth versus the total number of segments.
We split the available real-world data into a training set - 6 consecutive days of recordings; a validation set - 2 days; and a test set - 3 days recorded in a different month than training and validation data. The synthetic dataset replicates the distribution of events of the training set, and has therefore no statistical overlap with validation and test sets. As we are interested in evaluating the model performance on real-world data, we do not generate any synthetic data for testing.

We first train several models using only real-world ($\text{RW}$) data, increasing the training set size from $\SI{1}{\hour}$ to $\SI{144}{\hour}$ and training each model 4 times on randomly-chosen \lb{folds} within the available training data. Results, reported in Fig.~\ref{fig:res_train_real}, show the expected increasing trend in test accuracy when more training data is used. This confirms that the proposed model has the capacity to learn effectively from an increasing amount of data.
We then pre-train a model using $\SI{144}{\hour}$ of synthetic data generated as described in Sec.~\ref{subsec:data_generation} and fine-tune it with an increasing amount of real-world data ranging from $\SI{1}{\hour}$ to $\SI{144}{\hour}$. Real-world data for fine-tuning is randomly selected from the available real-world training data, repeating each training 4 times on different data subsets. Fig.~\ref{fig:acc_train_synth} shows the classification accuracy on the test set for the fine-tuned models. With no fine-tuning on real-world data ($\SI{0}{\hour}$) the model pre-trained only on synthetic data performs poorly, highlighting the need for a fine-tuning strategy that allows the model to adapt to the real data distribution. With as little as $\SI{6}{\hour}$ of real data for fine-tuning, however, the model is already capable of counting with an accuracy between $0.77$ and $0.87$ for all categories. The accuracy saturates when more than $\SI{24}{\hour}$ of real-world data are used for fine-tuning.

Fig.~\ref{fig:acc_comparison} compares accuracy between fine-tuned models with up to $\SI{144}{\hour}$ of real-world data and models trained from scratch on the same amount of real-world data. We hereby merge the directions of car and CV events and report, for each model, the average accuracy obtained in the 4 folds, together with accuracy ranges. The average accuracy per category obtained by fine-tuning on $\SI{24}{\hour}$ of real-world data ($\text{FT}_\text{24h}$) reaches $(0.88, 0.89, 0.95, 0.94)$ for (car-l2r, car-r2l, CV-l2r, CV-r2l) respectively, compared to $(0.63,0.63,0.86,0.87)$ obtained by training from scratch on real-world data only ($\text{RW}_\text{24h}$), with an average increase in accuracy of $0.25$ for cars and $0.08$ for CVs. A comparable accuracy requires \lb{between $\SI{72}{\hour}$ and} $\SI{144}{\hour}$ of real-world data. 
Fig.~\ref{fig:acc_comparison} also shows that fine-tuning with \lb{$\SI{12}{\hour}$ or more leads to a significant reduction in accuracy variation across folds, indicating that fine-tuning a model pre-conditioned to understand traffic is more stable than training from scratch on the same amount of data.} %

We finally evaluate the average miscount when the model predicts a wrong count. Here, we compute the mean absolute error on the misclassified samples only ($\text{MAE}_{\text{mis}}$), measuring by how much the prediction differs from the ground truth. It is worth noting that the lower bound for $\text{MAE}_{\text{mis}}$ is $1$. 
In Tab.~\ref{tab:comparison_table} we report both accuracy and $\text{MAE}_{\text{mis}}$ for each model trained from scratch ($\text{RW}$) and fine-tuned ($\text{FT}$). By comparing $\text{RW}_\text{24h}$ with $\text{FT}_\text{24h}$, we see that the increase in accuracy ($+0.25$ for cars, $+0.08$ for commercial vehicles) is accompanied by a significant reduction in $\text{MAE}_{\text{mis}}$ ($-0.10$ for both cars and commercial vehicles). This clearly shows the effectiveness of synthetic pre-training, allowing the $\text{FT}_\text{24h}$ to perform with high accuracy and low deviation in traffic counting.

\begin{table}[tb]
  \centering
  \begin{tabular}{cccccc}
    \toprule
    \multirow{2}{*}{} & \multicolumn{2}{c}{Accuracy} & \multicolumn{2}{c}{$\text{MAE}_{\text{mis}}$} & Real-world data\\
    \cmidrule(r){2-3}
    \cmidrule(r){4-5}
     & car  & CV  & car  & CV \\
    \midrule
    $\text{RW}_\text{144h}$ & 0.91 & 0.96 & 1.04 & 1.04 & $\SI{144}{h}$ \\
    $\text{RW}_\text{24h}$ & 0.63 & 0.86 & 1.16 & 1.13 & $\SI{24}{h}$ \\
    \midrule
    $\text{FT}_\text{24h}$ & 0.88 & 0.94 & 1.06 & 1.03 & $\SI{24}{h}$ \\
    $\text{FT}_\text{12h}$ & 0.81 & 0.90 & 1.10 & 1.08 & $\SI{12}{h}$\\
    $\text{FT}_\text{6h}$ & 0.77 & 0.87 & 1.13 & 1.12 & $\SI{6}{h}$ \\
    \bottomrule
  \end{tabular}
  \caption{Accuracy and mean absolute error on misclassified segments ($\text{MAE}_{\text{mis}}$) for models trained on real-world ($\text{RW}$) data only, or pre-trained on synthetic data and then fine-tuned ($\text{FT}$) on increasing amount of real-world data.}
  \label{tab:comparison_table}
  \vspace{-7pt}
\end{table}

\section{Conclusion and Future Work}
\label{sec:conclusions}
In this work we propose an AVC system based on a CRNN architecture trained on a mixture of synthetic and real-world data. The model relies on audio data captured by a 4-channel linear microphone array located at the side of the road. It is designed to count traffic for different vehicle types and direction of transit, and it is tested on real data from a two-lane road in moderate traffic density conditions. The proposed synthetic data generation procedure enables the model pre-training on simulated samples, with great benefit to reduce the amount of real-world data necessary to reach a high counting accuracy.

Future works will evaluate different data generation procedures to separately analyze the impact of the road/tire interaction and engine noises on the counting performance, as well as the generalization capabilities to different deployments. %

\bibliographystyle{IEEEbib}
\bibliography{bibliography}

\end{document}